\documentclass[sigconf]{acmart}

\AtBeginDocument{%
  \providecommand\BibTeX{{%
    \normalfont B\kern-0.5em{\scshape i\kern-0.25em b}\kern-0.8em\TeX}}}

\setcopyright{acmlicensed}
\copyrightyear{2025}
\acmYear{2025}
\acmDOI{XXXXXXX.XXXXXXX}

\usepackage{multirow}
\usepackage{subfig}
\usepackage{makecell}

\usepackage{algorithm}
\usepackage{algorithmic}

\usepackage{tikz}
\usetikzlibrary{arrows.meta,positioning}

\begin{document}

\title{Financial Wind Tunnel: A Retrieval-Augmented Market Simulator}

\author{Bokai Cao}
\authornote{Both authors contributed equally to this research.}
\affiliation{%
  \institution{The Hong Kong University of Science and Technology (Guangzhou)}
  \country{Guangzhou, China}
}
\affiliation{%
  \institution{IDEA Research, International Digital Economy Academy}
  \country{Shenzhen, China}
}
\email{mabkcao@connect.hkust-gz.edu.cn}

\author{Xueyuan Lin}
\authornotemark[1]
\affiliation{%
  \institution{The Hong Kong University of Science and Technology (Guangzhou)}
  \country{Guangzhou, China}
}
\affiliation{%
  \institution{IDEA Research, International Digital Economy Academy}
  \country{Shenzhen, China}
}
\email{xlin058@connect.hkust-gz.edu.cn}

\author{Yiyan Qi}
\affiliation{%
  \institution{IDEA Research, International Digital Economy Academy}
  \country{Shenzhen, China}
}
\email{qiyiyan@idea.edu.cn}

\author{Chengjin Xu}
\affiliation{%
  \institution{IDEA Research, International Digital Economy Academy}
  \country{Shenzhen, China}
}
\email{xuchengjin@idea.edu.cn}

\author{Cehao Yang}
\affiliation{%
  \institution{The Hong Kong University of Science and Technology (Guangzhou)}
  \country{Guangzhou, China}
}
\affiliation{%
  \institution{IDEA Research, International Digital Economy Academy}
  \country{Shenzhen, China}
}
\email{cyang289@connect.hkust-gz.edu.cn}

\author{Jian Guo}
\authornote{Corresponding Author.}
\affiliation{%
  \institution{IDEA Research, International Digital Economy
Academy
}
  \country{Shenzhen, China}
}
\email{guojian@idea.edu.cn}

\renewcommand{\shortauthors}{Bokai Cao, Xueyuan Lin, Yiyan Qi, Chengjin Xu, Cehao Yang and Jian Guo.}

\begin{abstract}

Market simulator tries to create high-quality synthetic financial data that mimics real-world market dynamics, which is crucial for model development and robust assessment. Despite continuous advancements in simulation methodologies, market fluctuations vary in terms of scale and sources, but existing frameworks often excel in only specific tasks. To address this challenge, we propose \textbf{Financial Wind Tunnel} (FWT), a retrieval-augmented market simulator designed to generate controllable, reasonable, and adaptable market dynamics for model testing. FWT offers a more comprehensive and systematic generative capability across different data frequencies. By leveraging a retrieval method to discover cross-sectional information as the augmented condition, our diffusion-based simulator seamlessly integrates both macro- and micro-level market patterns. Furthermore, our framework allows the simulation to be controlled with wide applicability, including causal generation through "what-if" prompts or unprecedented cross-market trend synthesis. Additionally, we develop an automated optimizer for downstream quantitative models, using stress testing of simulated scenarios via FWT to enhance returns while controlling risks. Experimental results demonstrate that our approach enables the generalizable and reliable market simulation, significantly improve the performance and adaptability of downstream models, particularly in highly complex and volatile market conditions. Our code and data sample is available at \url{https://anonymous.4open.science/r/fwt_-E852}
\end{abstract}



\maketitle

\section{Introduction}

The acceleration of market evolution and globalization has tightly interconnected financial systems at various scales, creating ripple effects and causing significant fluctuations. In the rapidly evolving field of quantitative finance, data-driven innovations and applications are still constrained by increasing market complexity and volatility. Quantitative models often struggle to navigate these intricate data dynamics, which can hinder their ability to consistently capture market movements. Their stability and adaptability are severely compromised when dealing with unprecedented trends and risks.

In order to improve the performance of investment models, the availability of large-scale high-quality data is essential. But accessing real-world financial data can be challenging due to limited size, privacy concerns, and cost barriers~\citep{kannan2024review}, particularly when acquiring structured price data. Against this backdrop, the concept of market simulator emerges, market simulator aims to generat simulation data with statistical authenticity by learning assets' characteristics and movements~\citep{cont2022tail, li2024mars, xia2024market, vuletic2024fin, zhang2022data}. Synthetic market dynamics can be used to augment real data, enabling models with more extensive incremental training, robustness assessment, decision optimization, and risk management. 

However, the development of a market simulator today faces numerous limitations. Firstly, patterns and fluctuation often persist across different data frequencies~\citep{guo2024large}. Simulation across different scales is influenced by distinct market behaviors and characteristics, leading to significant differences in distribution and structure. Existing market simulators typically predict only at specific frequencies. There is a need for a versatile framework to model and link market conditions at different scales, ranging from macro- to micro-levels. Secondly, market dynamics are increasingly influenced by external capital and other markets~\citep{SARWAR20141}, Similar to the formulation of Barra risk factors~\citep{lu2019barra, rosenberg1976common}, risks can often be identified through cross-market pattern learning, especially in markets with longer histories or those impacted by specific events. However, there is still a lack of simulators capable of mimicking the interaction between cross-market movements. Ultimately, extreme market conditions bring challenges to stability while creating trading opportunities. A market simulator needs causal generation capabilities to generate controllable market fluctuations based on pre-defined assumptions of risky conditions.

\begin{figure*}
    \centering
    \includegraphics[width=\linewidth]{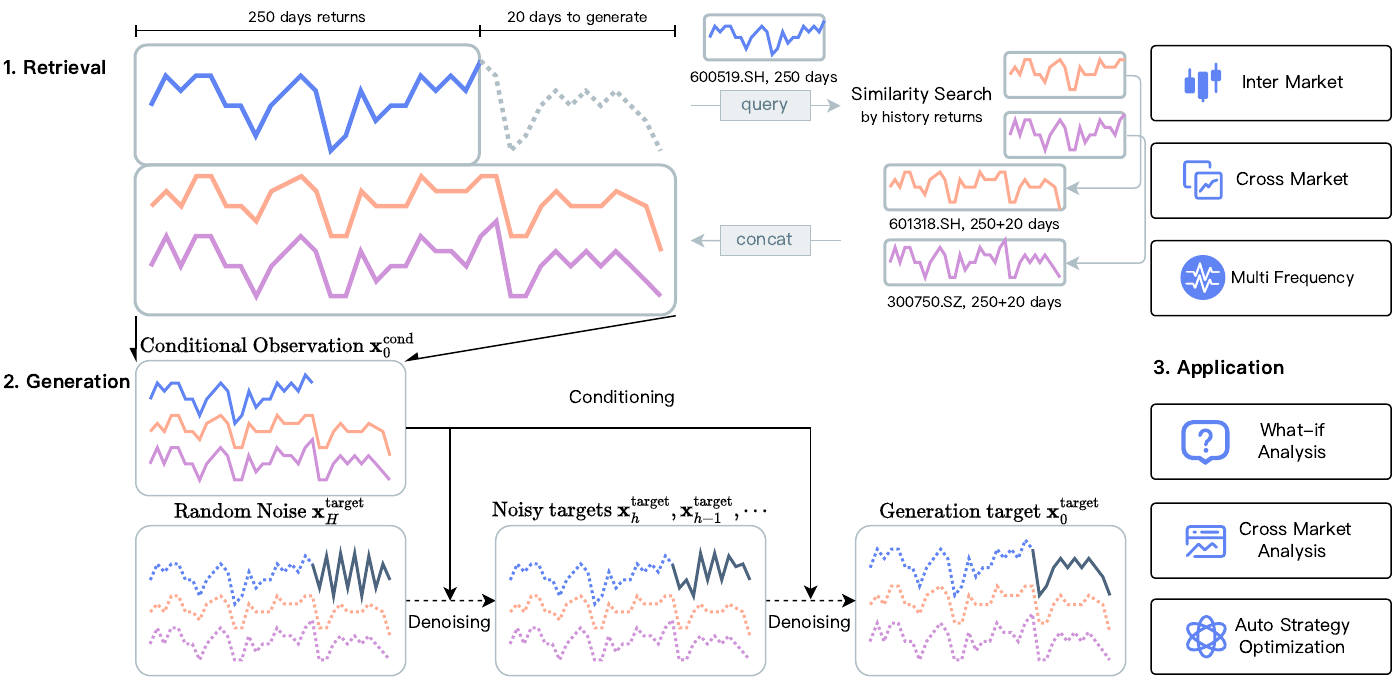}
    \caption{The overall procedure of Financial Wind Tunnel (FWT). There are three core modules: retrieval, generation, and application. The data sources for retrieval include inter-market stocks, cross-market stocks, and multi-frequency dynamics. With retrieved series as conditional observation, FWT generates new time series for downstream tasks like what-if analysis, cross-market analysis, and trading strategy optimization.}
    \label{fig:main}
\end{figure*}

To resolve the above limitations, we propose Financial Wind Tunnel (FWT) in this paper. FWT is a diffusion-based market simulator that retrieves cross-sectional information to augment financial time-series generation. Inspired by the wind tunnel experiments in aerospace engineering, it aims to construct a highly reliable and diverse simulation environment for downstream financial models. Motivated by the latest progress in diffusion-based generative models that incorporate guidance mechanisms~\citep{ho20,ho21,rombach22}, the problem could potentially be resolved through the utilization of a conditional diffusion model, since a simulator is essentially focused on generation tasks rather than prediction. By employing a novel approach to retrieve the most relevant stocks, FWT adapts augmented generation across various frequencies. It can also achieve simulation customization through different retrieval rules, and enable "what-if" generation and cross-market pattern learning. In this way, we can preserve the intrinsic asset property and market characteristics, while also mimicking the desired trends.

Based on that, we develop an downstream model optimizer that leverages the simulated market scenarios provided by the financial wind tunnel to test and dynamically adjust investment models, achieving the dual goals of maximizing returns and controlling risks. Through iterative wind tunnel testing cycles, adjusting parameters, and analyzing outcomes, models can be optimized for efficiency, stability, and robustness.

In summary, the contributions of this paper are as follows:
\begin{itemize}
    \item We propose a diffusion-based rtrieval-augmented market simulator, Financial Wind Tunnel (FWT), which generates reliable, controllable and robust simulation across various data frequencies. We introduce an innovative return-correlation retrieval strategy to augment the conditional diffusion generation, providing strong support for synthesis and effectively evaluating its validity. 
    \item We develop and validate that FWT maintains strong generation performance in "what-if" type causal generation tasks by modifying the expectations and prompt for future market conditions. Meanwhile, we introduced a cross-market mechanism to simulate the characteristic fluctuations of other markets.
    \item We conduct comprehensive evaluations on extensive datasets, demonstrating the generation quality and simulation-based automated optimization can significantly enhance downstream investment strategies, improving the stability and adaptability of traditional quantitative models in complex and volatile market environments. This provides new perspectives and methods for advancing the field of quantitative financial investment.
\end{itemize}

\section{Related Works}
\label{sec:related_works}

Financial market simulation aims to synthesize realistic quotes of financial markets. There are four main categories: rule-based, time-series based, order flow-based, and multi-agent-based.
\paragraph{Rule-based} 
Rule-based approaches figured prominently in early market simulation studies. These methods rely on predefined rules to emulate market participants' behavior, such as executing trades under specific price conditions based on assumptions like trend following or mean reversion~\citep{raberto01,palmer94}, or simplistic resample methods like bootstrapping or random sampling. While simple, straightforward and interpretable, rule-based models often lack the flexibility to adapt to the dynamic nature of real-world markets with non-linear relationships and complex dependencies~\citep{kannan2024review}. Their fixed rules fail to capture the interplay and evolution of market factors, limiting their applicability in simulating diverse modern market trends.

\paragraph{Time Series-based}
Time series-based approaches leverage historical price and volume data to simulate market dynamics. These methods often employ stochastic models to generate synthetic data that replicates the statistical properties of real markets. Traditional time-series model based on GARCH and ARIMA studies price forecasting and volatility estimation to synthesize data~\citep{bauwens2006multivariate, coletta21,li20}.  The emergence and advancement of neural network and deep learning have led to a fundamental change in the way synthetic data is generated. Generative models like Variational Autoencoders (VAEs)~\citep{rezende2014stochastic, kingma2013auto}, Generative Adversarial Networks (GANs)~\citep{fu2019time, zhang2022data, goodfellow2014generative}, diffusion models~\citep{huang2024generative} and prediction models include LSTM~\citep{zhou2018stock}, Transformer ~\citep{vaswani2017attention} provide an advanced approach for capturing the complex patterns and structures present in financial data. Base on these models, we can directly predict market trajectories based on historical data~\citep{coletta22,coletta23}, generated market fluctuations and market risks~\citep{cont2022tail, xia2024market, vuletic2024fin, wiese2020quant}. Although effective in preserving data characteristics, these methods are limited in capturing the causal interactions between market participants and struggle to capture extreme condition, the direction of data generation is also difficult to control effectively.


\paragraph{Order Flow-based} 
With the in-depth study of market microstructure and computational efficiency improvement, simulation methods based on order flow are emerging. These approaches model the dynamic process of order arrivals, executions, cancellations and limit order book formation to simulate price formation and volatility~\citep{chiarella02,chiarella09,li2020generating}. By treating markets as arenas for order interactions, these models aim to uncover the mechanisms behind price dynamics. Examples include frameworks such as DeepLOB\citep{zhang2019deeplob}, MarS \cite{li2024mars}, DiGA~\citep{DiGA}. However, the challenges lie in the complexity of accurately modeling real-world order data and the computational overhead of processing large-scale order flows. At the same time, this methodology can only focus on short-term market behavior and high-frequency price fluctuations, performing poorly on larger economic cycles.

\paragraph{Multi-Agent-based} 
In recent years, market simulation methods based on multiple agents have received a lot of attention.
This paradigm models traders, investors, and dealers as agents with distinct objectives and decision-making strategies~\citep{lux99,abides20}. Agents interact with each other and the market environment, leading to emergent phenomena that resemble real-world markets. Such methods excel in capturing market heterogeneity and complexity. However, they face challenges in balancing the realism and computational efficiency of agent behaviors and in calibrating model parameters to ensure accuracy~\citep{wang17} in long-term sequence generation.

\section{Method}
\label{sec:method}
In this section, we propose the method and structure of Financial Wind Tunnel (FWT). First, we describe the retrieval augmentation method in detail in Section~\ref{sec:method:retrieval}, showing how we construct context signals for conditional generation and enables controllable simulation. Then, we illustrate how to retrieval-augmented generate market dynamics with conditional diffusion model via masked modeling in Section~\ref{sec:method:gen}. Finally, we construct an automatic strategy optimizer in Section~\ref{sec:method:opt}, showcasing the usage of our market simulator.

\begin{figure*}[th]
  \centering
  \includegraphics[width=.8\linewidth]{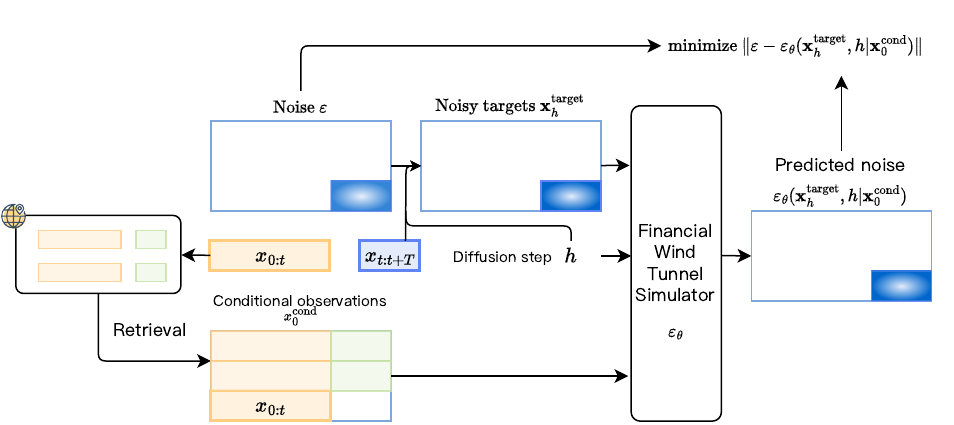}
  \vspace {-1em}
  \caption{The self-supervised training procedure of FWT. FWT synthesizes $x_{t:t+T}$ when given a history time series $x_{0:t}$, which is used to query top-$k$ most relevant stocks to construct conditional observations. FWT predicts noise from noisy target $\mathbf{x}_h^{\text{target}}$ at the diffusion step $h$ to recover $x_{t:t+T}$.}
  \label{fig:method}
\end{figure*}

\subsection{Retrieval}
\label{sec:method:retrieval}

In order to enhance the credibility of the generated series and the target series, we perform retrieval-augmentation to construct reliable context signals cross-sectionally.
We consider the returns series of the target stock $S_0$ to be split into historical part $X_{S_0}$ over time interval $[0, t)$ and future part $Y_{S_0}$ on $[t, t+T)$ as generation target, denoted as:
\begin{equation}
X_{S_0} = \{r_{0, 1:t}\} \in \mathbb{R}^t, \text{and } Y_{S_0}=\{r_{0, t:t+T}\} \in \mathbb{R}^T
\end{equation}
where returns $r_{i}=\frac{p_i-p_{i-1}}{p_{i-1}}$ are derived from price $p_i$.
Using a similarity measure, such as Dynamic Time Warping (DTW) ~\citep{yadav2018dynamic} or Pearson Correlation of stock return, we retrieve the top-$K$ stocks whose historical price movements during $[0, t)$ are most similar to $S_0$. Let these top-$K$ stocks be denoted as:
\begin{equation}
\begin{aligned}
\text{SimilarStocks}(S_0, 0:t, K) = \{S_1, S_2, \dots, S_K\} \\
X_{S_i} = \{r_{i, 1:t}\} \in \mathbb{R}^t, \; Y_{S_i} = \{r_{i, t:(t+T)}\} \in \mathbb{R}^T
\end{aligned}
\end{equation}
where $X_{S_i}$ is the historical return series of the $i$-th similar stock $S_i$, and $Y_{S_i}$ is its future return movements over the time interval $[t, t+T)$.
It should be noted that these stocks can be retrieved either within the same market or across different markets. 
By concatenation, the value of the observed multivariate time series is:
\begin{equation}
    \mathbf{X} = \{r_{0:K+1,0:t+T}\} \in \mathbb{R}^{(k+1)\times (t+T)}
\end{equation}
We also denote an observation mask as $\mathbf{M}=\{m_{0:K+1,0:t+T}\}\in \{0,1\}^{(k+1)\times (t+T)}$, where $m_{0,t:t+T}=0$ and $m_{\text{else}}=1$. The mask filters the future part $Y_{S_0}$ of stock $S_0$. Our goal is to synthesize a time series $Y'_{S_0}$ to approximate the real $Y_{S_0}$. To achieve the goal, we utilize the retrieval results as condition and deploy a diffusion method as is described in the next section\ref{sec:method:gen}.

We emphasize that the retrieval process can run on all frequencies of stock dynamics. This setting is named Multi-frequency Generation, whose result will be shown in Section~\ref{sec:exp:results}.
Besides, we can apply retrieval filtering rules to identify different forms of conditions to control the generation. in order to retrieve-generate data from various sources of time-series for downstream tasks.
In this paper, we explore two more cases: 1) Cross-market generation, which retrieves from other markets; and 2) What-if generation, which recalls from the provided time series database. The results will be presented in Section~\ref{sec:exp:results}.

\subsection{Generation}
\label{sec:method:gen}

We utilize a conditional diffusion model to generate the future price trajectory of the target stock $S_0$.
To begin with, in conventional DDPM~\cite{ddpm}, given data $\mathbf{x}_0 \sim p(\mathbf{x}_0)$, the forward process is a Markov-Gaussian process that gradually adds noise to obtain a perturbed sequence $\{x_1, x_2,\cdots,x_H\}$,
\begin{equation}
    q(\mathbf{x}_1,\cdots,\mathbf{x}_H\mid\mathbf{x}_0) = \prod_{h=1}^H q(\mathbf{x}_h\mid\mathbf{x}_{h-1}),
\end{equation}
where $q(\mathbf{x}_h\mid\mathbf{x}_{h-1}) = \mathcal{N}(\mathbf{x}_h; \sqrt{1-\beta_h} \mathbf{x}_{h-1}, \beta_h \mathbf{I})$, $\mathbf{I}$ is the identity matrix, $\beta_h$ is a learnable parameter, $H$ is the number of diffusion steps, $q$ represents the forward process, and $\mathcal{N}$ denotes the Gaussian distribution parameterized by hyperparameters $\{\beta_h\}_{h\in [H]}$. Perturbed samples are sampled via $x_h = \sqrt{\bar{\alpha}_h}x_{0} + \sqrt{1-\bar{\alpha}_h}\epsilon$, where $\epsilon \sim \mathcal{N}(0, \mathbf{I})$, $\alpha_h = 1-\beta_h$ and $\bar{\alpha_h} = \prod_{s=1}^h \alpha_s$.
For the reverse process, the model learns to predict noise $\epsilon$ from the noisy target $\mathbf{x}_h$ at the diffusion step $h$ to recover the original data $\mathbf{x}_0$.
It introduces a specific parameterization of $p_{\theta}(\mathbf{x}_{h-1}\mid\mathbf{x}_h)$:
\begin{equation}
\begin{aligned}
\mathbf{\mu}^{\text{DDPM}}(\mathbf{x}_h, h) &= \frac{1}{\alpha_h} (\mathbf{x}_h - \frac{\beta_h}{\sqrt{1-\alpha_h}}\mathbf{\epsilon}_\theta(\mathbf{x}_h, h)) \\
\mathbf{\sigma}^{\text{DDPM}}(\mathbf{x}_h, h) &= \sqrt{\bar{\beta}_h}
\end{aligned}
\end{equation}
The training loss~\cite{ddpm} for diffusion models is to predict the normalized noise $\epsilon$ by solving the following optimization problem:
\begin{equation}
    \min_{\theta} \mathbb{E}_{\mathbf{x}_0\sim p(\mathbf{x}_0), \mathbf{\epsilon}\sim \mathcal{N}(0, \mathbf{I}), h} \left[\left\|\mathbf{\epsilon} - \mathbf{\epsilon}_\theta(\mathbf{x}_h, h)\right\|_2^2\right],
\end{equation}
The denoising function $\mathbf{\epsilon}_\theta$ estimates the noise $\epsilon$ that was added to the noisy target $\mathbf{x}_h$ at the diffusion step $h$.

Then, we consider extending the parameterization of DDPM to the conditional case. We define a conditional denoising function $\mathbf{\epsilon}_\theta: (\mathcal{X}^{\mathrm{target}}\times\mathbb{R}\mid\mathcal{X}^{\mathrm{cond}})\to\mathcal{X}^{\mathrm{target}}$, which takes conditional observations $\mathbf{x}_0^{\mathrm{cond}}$ as inputs. We consider the following parameterization with $\mathbf{\epsilon}_\theta$:
\begin{equation}
\begin{aligned}
\mathbf{\mu}_\theta(\mathbf{x}_h^{\mathrm{target}}, h \mid \mathbf{x}_0^{\mathrm{cond}}) &= \mathbf{\mu}^{\text{DDPM}}(\mathbf{x}_h^{\mathrm{target}}, h, \mathbf{\epsilon}_\theta(\mathbf{x}_h^{\mathrm{target}}, h \mid \mathbf{x}_0^{\mathrm{cond}})) \\
\mathbf{\sigma}_\theta(\mathbf{x}_h^{\mathrm{target}}, h \mid \mathbf{x}_0^{\mathrm{cond}}) &= \mathbf{\sigma}^{\text{DDPM}}(\mathbf{x}_h^{\mathrm{target}}, h)
\end{aligned}
\label{eq:mu}
\end{equation}
For the sampling, we set all observed values $\mathbf{x}_0$ as conditional observations $\mathbf{x}_0^{\mathrm{cond}}$ and all missing values as the target $\mathbf{x}_0^{\mathrm{target}}$ to be generated. We sample noisy targets $\mathbf{x}_h^{\mathrm{target}} = \sqrt{\bar{\alpha}_h}\mathbf{x}_0^{\mathrm{target}} + \sqrt{1-\bar{\alpha}_h}\epsilon$ and train $\mathbf{\epsilon}_\theta$ by minimizing the loss:
\begin{equation}
\min_{\theta} \mathbb{E}_{\mathbf{x}_0\sim p(\mathbf{x}_0), \mathbf{\epsilon}\sim \mathcal{N}(0, \mathbf{I}), h} \left[\left\|\mathbf{\epsilon} - \mathbf{\epsilon}_\theta(\mathbf{x}_h^{\mathrm{target}}, h \mid \mathbf{x}_0^{\mathrm{cond}})\right\|_2^2\right].
\end{equation}
As is illustrated in Figure~\ref{fig:method}, given a sample $\mathbf{x}_0$, we separate it into conditional observations $\mathbf{x}_0^{\mathrm{cond}}$ and the target $\mathbf{x}_0^{\mathrm{target}}$. In market synthesization, the target $\mathbf{x}_0^{\mathrm{target}}=\mathbf{M}*\mathbf{X}$ is the future price trajectory of the target stock $S_0$ over the interval $[t, t+T)$, and the conditional observations $\mathbf{x}_0^{\mathrm{cond}}=(1-\mathbf{M})*\mathbf{X}$, where $*$ is element-wise multiplulation. Under such settings, the model $\mathbf{\epsilon}_{\theta}$ has the same shapes of input and output tensors.
The diffusion objective naturally becomes a self-supervised training procedure inspired by masked language modeling~\cite{kenton2019bert}. We utilize Transformer as the backbone for diffusion. More implementation details of $\mathbf{\epsilon}_{\theta}$ can be found in Section~\ref{sec:exp:settings}. We also provide the training procedure of Financial Wind Tunnel in Algorithm~\ref{alg:training} and the generation procedure in Algorithm~\ref{alg:sampling}.

\begin{algorithm}[t]
\caption{Training of Financial Wind Tunnel}
\label{alg:training}
\begin{algorithmic}[1]
\STATE {\bfseries Input:} Training data distribution $q(\mathbf{X})$, observation mask $\mathbf{M}$, number of iterations $N_{\rm iter}$, noise schedule $\{\alpha_h\}_{h\in[H]}$
\STATE {\bfseries Output:} Trained denoising function $\epsilon_\theta$
\FOR{$i=1$ {\bfseries to} $N_{\rm iter}$}
\STATE $h \sim \textrm{Uniform}(\{1,\ldots,H\})$, $\mathbf{X} \sim q(\mathbf{X})$ 
\STATE Separate $\mathbf{X}$ into conditions $\mathbf{x}_0^{\rm cond} = (1-\mathbf{M}) \odot \mathbf{X}$ and targets $\mathbf{x}_0^{\rm target} = \mathbf{M} \odot \mathbf{X}$
\STATE $\epsilon \sim \mathcal{N}(\mathbf{0},\mathbf{I})$ with same dimension as $\mathbf{x}_0^{\rm target}$
\STATE Compute noisy targets $\mathbf{x}_h^{\rm target} = \sqrt{\bar{\alpha}_h}\mathbf{x}_0^{\rm target} + \sqrt{1-\bar{\alpha}_h}\epsilon$ \hfill $\triangleright$ Eq.(4)
\STATE Update $\theta$ via $\nabla_\theta \left\| \epsilon - \epsilon_\theta(\mathbf{x}_h^{\rm target}, h \mid \mathbf{x}_0^{\rm cond}) \right\|_2^2$ \hfill $\triangleright$ Eq.(8)
\ENDFOR
\end{algorithmic}
\end{algorithm}

\begin{algorithm}[htbp]
   \caption{Generation with Financial Wind Tunnel}
   \label{alg:sampling}
\begin{algorithmic}[1]
   \STATE {\bfseries Input:} Observed data $\mathbf{X}$, mask $\mathbf{M}$, trained $\epsilon_\theta$
   \STATE {\bfseries Output:} Generated trajectory $\mathbf{x}_0^{\rm target}$
   \STATE Initialize $\mathbf{x}_H^{\rm target} \sim \mathcal{N}(\mathbf{0},\mathbf{I})$ matching $\mathbf{M}$'s dimension
   \FOR{$h=H$ {\bfseries to} $1$}
   \STATE Compute $\mu_\theta(\mathbf{x}_h^{\rm target}, h \mid \mathbf{x}_0^{\rm cond})$ via Eq.(7)
   \STATE Sample $\mathbf{x}_{h-1}^{\rm target} \sim \mathcal{N}(\mu_\theta, \bar{\beta}_h\mathbf{I})$
   \ENDFOR
\end{algorithmic}
\end{algorithm}

\subsection{Automatic Strategy Optimizer}
\label{sec:method:opt}

Financial wind tunnel can provide an environment for market simulation and risk imitation, which test and enhance the performance of downstream model prediction, portfolio optimization, and risk control, especially in facing extreme market conditions. Synthetic data offers numerous possibilities for enhancing current machine learning-based quantitative investment models rather than limited-scale real market data. To demonstrate the utility of it, we develop an automatic strategy optimizer that refines trading strategies based on our FWT. The optimizer operates in two modes:

\begin{itemize}
    
    \item \textbf{Model-Based Optimization:} Optimizer with simulation data automatically optimize hyperparameter tuning, feature engineering, model architecture search, and ensemble methods of downstream models. Improves the ability of the model to generalize across different market trends and resist overfitting. The model will also be trained on data with controlled generation, such as large-volatility, to improve prediction accuracy under significant market trends.
    
    \item \textbf{Rule-Based Optimization:} This mode focuses on filtering for predefined rule-based portfolio metrics, such as returns, drawdown, and Sharpe ratio. The optimizer systematically evaluates the performance of different portfolio configurations and eliminates those that underperform during extreme market conditions.

\end{itemize}

Through these functionalities, the automatic strategy optimizer demonstrates how the Financial Wind Tunnel facilitates downstream strategy development and testing in a controlled, repeatable environment.

\subsection{Evaluation}

As financial task acting as a "wind tunnel", our evaluation is primarily divided into two parts: First, the quality of the generated data, which must both preserve its inherent statistical properties and effectively learn to mimic subsequent market trends. Second, whether the generated data can enhance the performance of downstream models when applied. 

Regarding generation quality, we validate whether the simulation closely resembles its real-world trends by assessing real-simulation correlation and its cross-sectional market ranking. Mathematically, the former one is typically defined as the Pearson correlation coefficient between the generated data $\hat{r}_i$ and the actual price movement $r_i$ over a specific time horizon: $\frac{\textit{Cov}(r_i, \hat{r}_i)}{\sigma_{r_i} \sigma_{\hat{r}_i}}$, where \(\textit{Cov}\) is the covariance and \(\sigma\) represents the standard deviation. Introducing the market ranking of correlation between real and simulation is one of our novel contributions. Considering that our task is generation rather than prediction, we need to determine whether the synthetic data, while replicating the original trend, can outperform other real market data. It compares and ranks $Coef(r_i, \hat{r}_i)$ with $Coef(r_i, r_j)  ( \small j \in \{1, 2, \dots, n\})$, where $r_j$ represents the real data of any other cross-sectional stock over the retrieval period.

\begin{equation*}
    R_{\textit{Coef}(r_i, \hat{r}_i)} = \operatorname{rank}\left( \textit{Coef}(r_i, \hat{r}_i), \textit{Coef}(r_i, \{\hat{r}_i, r_1, r_2, \dots, r_n\}) \right)
\end{equation*}

It not only validates whether the generated data retains the original trend and statistical properties of the target stock but also assesses if it aligns with the overall future market style and outperform most similar stocks. Beyond absolute similarity, this approach introduces a relative comparison and enables quantitative evaluation for cross-market and what-if generation.

In terms of downstream tasks, we primarily focus on the investment metrics including annualized returns, max drawdown, Sharpe ratio of the quantitative model before and after FWT enhancement, and provide a more intuitive comparison using the return curve. The annualized return is the annualized growth of the net value, max drawdown measures the maximum loss from a peak to a trough during a specific period and Sharpe ratio is the ratio of excess return to the standard deviation of returns, evaluating the risk-adjusted return.

\section{Experiments}
\label{sec:exp}
In this section, we present the experiment settings and results on real-world datasets to evaluate the quality of data generated by FWT as well as its model optimization capability.

\subsection{Experiment Settings} \label{sec:exp:settings}

To demonstrate the model performance and generalization ability of our market simulator, we systematically conduct experiments in the following four aspects.

\begin{itemize}
    \item Multi-frequency Generation: We will train and evaluate the model across different financial data frequencies to demonstrate its strong and stable representational capability at various scales. We will also demonstrate the effectiveness of the model in cross-frequency transfer learning.
    \item Cross-market Generation: We will validate the data generation quality in cross-market scenarios, focusing on the model's ability to learn and generate data from different market patterns while preserving the original trends and statistical properties. 
    \item What-if Generation: We will simulate various financial scenarios and use a "what-if" approach to assess the generators' causal generation ability when simulate different predefined future market conditions, especially extreme fluctuations.  
    \item Model Enhancement: We will apply the generated data from the generator to downstream tasks. This will demonstrate the generator's role as a financial wind tunnel for simulating and enhancing the stability and performance of market models in the face of changing market environment.
    
\end{itemize}

\begin{table}[t]
\caption{Dataset statistics of different data frequencies.}
\vspace{-2mm}
\resizebox{\columnwidth}{!}{
\begin{tabular}{lccccc}
\toprule
Freq & Train & Valid & Test & Samples \\
\midrule
1Week   & 2010.01-2018.12  & 2019.01-2020.12  & 2021.01-2024.12 & 220K  \\
1Day    & 2010.01-2018.12  & 2019.01-2020.12  & 2021.01-2024.12 & 1M  \\
1Hour   & 2010.01-2018.12  & 2019.01-2020.12  & 2021.01-2024.12 & 4M   \\
1Min   & 2018.01-2021.12  & 2022.01-2022.12  & 2023.01-2024.12 & 125M   \\
1Tick   &  2024.01-2024.06  & 2024.07-2024.09  & 2024.10-2024.12 & 326M   \\
\bottomrule
\end{tabular}\label{tab: dataset}}
\end{table}
  
\paragraph{Datasets} 
\begin{figure*}[th]
  \centering
  \includegraphics[width=0.7\linewidth]{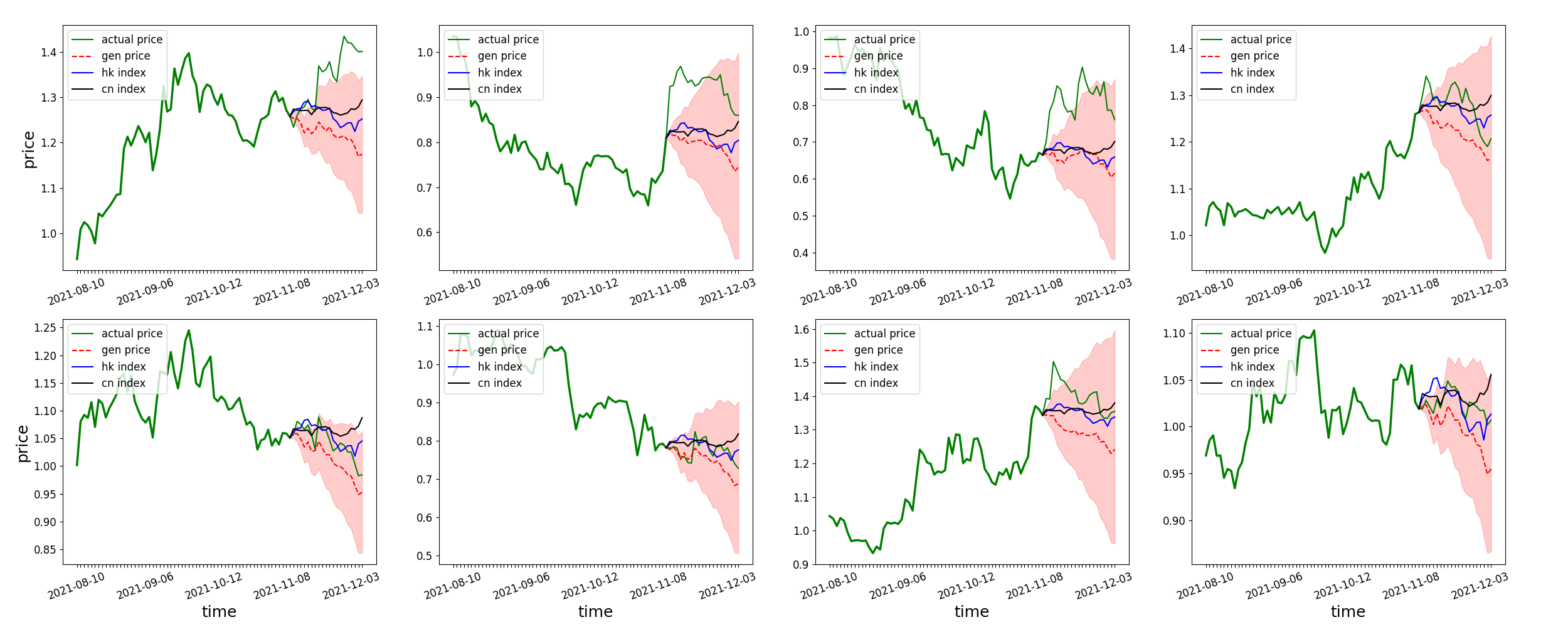}
  \vspace {-1em}
  \caption{Cross-market generation case study: Liquidity crisis in the Hong Kong stock market.}
  \label{fig:case}
\end{figure*}

To conduct a comprehensive experiment and evaluation of FWT, we primally choose CSI300 stocks in the Chinese stock market. CSI300 consists the 300 largest and most liquid stocks listed on the Shanghai and Shenzhen stock exchanges and is considered to be the most representative set of stock securities in China. We use stock price data from various frequencies spanning over from 2010 to 2024. All the data were split into training, validation, test sand the data is standardized and outlier-handled, dataset statistics and configurations are summarized in Table ~\ref{tab: dataset}. Additionally, in our cross-market study, we have utilize relevant data from the Hong Kong Stock Exchange (HKSE) with a aligned time period.

\paragraph{Implementation Details}
The experiment is conducted on all component stocks of CSI300 for individual market experiments and HKSE also used for cross-market one. Hyperparameters for FWT are same across all experiments, the model's history window length is 250 trading days to retrieve highest 16 stocks with excess return correlation. The generated length is 20. The diffusion generate 100 steps and we leverage a transformer of 4 layers and 8 attention heads. We apply Adam optimizer with an initial learning rate of 1.5e-4 to train 500 batches. All the experiments are conducted by NVIDIA 4090Ti GPUs.

For the downstream task, the foundation quantitative model is a Transformer-based for long-term forecasting with daily frequency. Data utilized for training ranges from 2018 to 2021, including both real data and generated market simulation, with 2022 as the validation and 2023 as the test. Model are also based on CSI300 constituent stocks, predicts over 20 days while long-short and rebalance 1/4 position of the portfolio every 5 days.

\subsection{Main Results}
\label{sec:exp:results}

\paragraph{Multi-frequency Generation} 
We conducted experiments using various data frequencies as Table ~\ref{tab:freq_res} to validate the generative capability of FWT across different scales. The result shows that, whether macro or micro market trends, FWT effectively captures and simulates the market patterns since the indicators of different frequencies are both high. 

Among these scales, the daily and hourly levels achieved the best synthesis results. The correlation of simulation and its real market trend reached over 0.6, indicating a good fit to the data distribution and statistical features of the original stock. Furthermore, the market ranking shows that the synthetic data outperforms most other actual stock trends in terms of the correlation with the future real trends, demonstrating that the generated data, while preserving its intrinsic properties, effectively adapts to future market dynamics. The weekly level, due to fewer training samples and a longer prediction period (about five months), exhibited some instability. Tick-level data, on the other hand, is highly disturbed by short-term market behaviors, leading to relatively lower correlation. However, it still achieved above 0.4, which is considered quite high in the financial domain.

\begin{table}[t]
\centering
\caption{Different frequency generation quality.}
\vspace{-2mm}
\setlength{\tabcolsep}{1mm}{
\begin{tabular}{lcccc}
\toprule
Freq &  Market Ranking & Correlation \\
\midrule
1Week   &  95.78\%  & 0.369   \\
1Day   &  99.71\%  & 0.645  \\
1Hour  & 99.29\%  & 0.602   \\
1Min   & 97.69\%  & 0.566   \\
1Tick   &   85.43\% & 0.491  \\
\bottomrule
\end{tabular}\label{tab:freq_res}}
\end{table}

\begin{table}[t]
\centering
\caption{Cross frequency transfer learning results.}
\vspace{-2mm}
\resizebox{\columnwidth}{!}{
\begin{tabular}{lcccc}
\toprule
\small Freq & \small Base Model & \small Transfer learning & \small Market Ranking & \small Correlation \\
\midrule
1Week & 1Week & No&95.78\%  & 0.369   \\
1Week & 1Hour  &  No&93.54\%  & 0.401   \\
1Week & 1Hour   & Yes& 96.88\%  & 0.436   \\
\bottomrule
\end{tabular}\label{table:transfer}}
\end{table}

Furthermore, enabled by our generalized retrieval-augmented conditional generation architecture, we observed that the model exhibits transfer capabilities across different frequencies. Specifically, for the weekly generation application scenario with limited training samples, we designed comparative experiments to a base model which first trained on hourly data and further transfer to weekly application scenario. We compared its performance before and after transfer learning against the original weekly model. The results demonstrate that: (1) Similar patterns exist across different frequencies and can be leveraged. Since the base models from other frequencies can zero-shot simulate and perform well without incremental training.(2) Transfer learning across frequencies helps further improvement of model performance. Models trained on larger training sets demonstrate better generation performance after fine-tuning.

  \begin{figure*}[th]
  \centering
  \includegraphics[width=.9\linewidth]{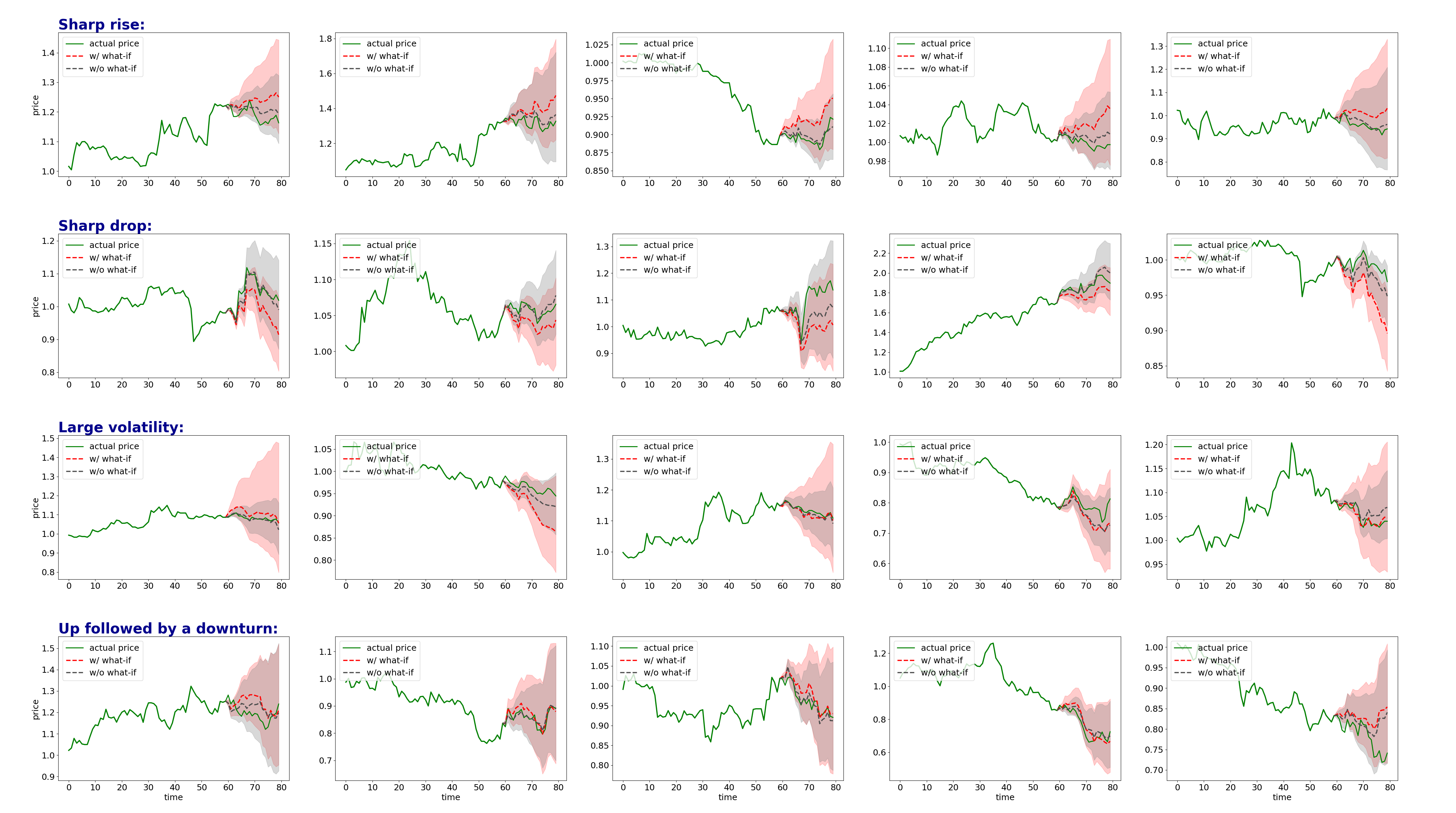}
  \vspace {-1em}
  \caption{Cases of what-if generation. The figure illustrates a comparison between actual trends (green) with the results generated with and without what-if condition (red and grey respectively) under four different prompts.}
  \label{fig:whatif}
\end{figure*}

\paragraph{Cross-market Generation} 
An important characteristic of FWT is its ability to simulate previously unseen fluctuations by simulate cross-market trends and risks. In this experiment, we enable stocks in the Chinese stock market to learn the patterns of the Hong Kong stock market. We retrieve the historical data from HKSE by excess return correlation with a specific CSI300 stock. We limit the scope of the retrieval to the HKSE, using the generation time period return of stocks with the most similar historical trends in the HKSE as the condition to augment generation.

To assess whether the generated data retain the original market characteristics while successfully simulating other markets trends, we calculate the IC and its market ranking between the generated sequences and the original A-share sequences across Hong Kong stock market. The results indicate that the IC reached an absolute value of 0.475, suggesting that the cross-market condition retains its historical attributes. Meanwhile, the market ranking achieved 91.83\%, outperforming the direct fitting of the majority of Hong Kong stocks.

We have included a case study to demonstrate the cross-market capability of FWT. At the end of 2021, under the influence of tightening international monetary policies, external capital outflows, and repeated defaults by real estate companies, the Hong Kong stock market experienced much larger drawdowns and volatility compared to the A-share market. During this period, we were able to use A-share data to specifically learn the trends of the Hong Kong stock market, preparing for potential liquidity crises in the future. As shown in the Figure\ref{fig:case}, while synthetic and real data still exhibit a strong correlation, the price series has also largely learned the volatility patterns of the Hong Kong stock market.

\paragraph{What-if Generation} 

FWT, as a market simulator, needs to maintain good generation efficiency under high-volatility and high-risk market environments, while also possessing the ability to actively control the generation direction. Based on the rule-based retrieval generation process to filter out market data that matches the input prompt as our retrieval space, we propose a what-if generation approach to construct the intended environment. 

The generated result is shown in the Figure\ref{fig:whatif}. The figure illustrates the controllable generation performance under four typical what-if conditions, comparing the real market trend with the generation w/ and w/o prompts, and the generation with prompts. The green line represents the actual historical trends and future market data, while the gray dashed line indicates normal generation and the red dashed line represents the what-if generation. The shaded regions in gray and red represent the generation intervals  at the 25\% and 75\%  quantiles, respectively. It can be observed that both generated patterns closely mimic the real data while what-if generation successfully simulating the given what-if prompt, showing correct direction shifts in the overall trend.

\paragraph{Model Enhancement} 
FWT has the ability to provide a testing environment for downstream models, enabling simulation, training, and risk assessment. We take a transformer-based long-horizon forecasting quantitative investment model as an example to demonstrate the optimization process of FWT, seeing implementation details as \ref{sec:exp:settings}. We compare the model's performance when trained solely on real market data versus when trained with additional FWT-generated synthetic data in handling out-of-sample market environment.

\begin{figure}[th]
  \centering
  \includegraphics[width=.8\linewidth]{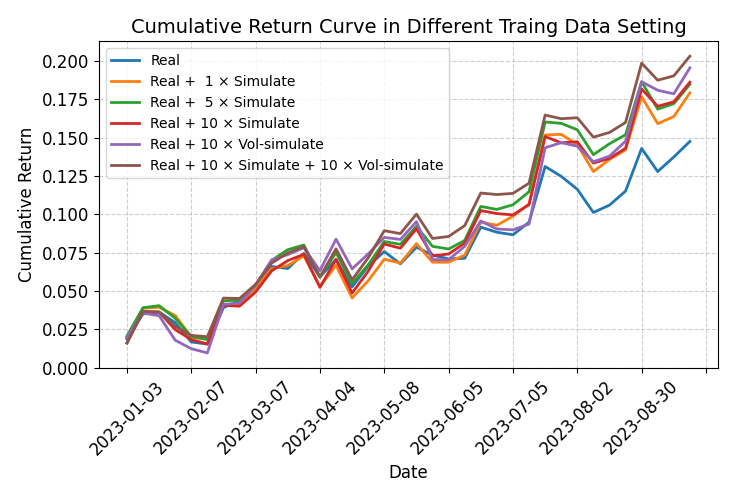}
  \vspace {-1em}
  \caption{Cumulative return curve of hedge portfolio training from different dataset}
  \label{fig:ds_task}
\end{figure}

Based on the size of real market dataset, we introduce additional simulated data at scales of 1x, 5x, and 10x, as well as 10x simulated data. Since high-volatility market environments are most suitable for quantitative models to identify arbitrage opportunities, while also being the most prone to risks and drawdowns, making their management particularly crucial during execution. We also set up a group of volitility-simulation, controlled via what-if generation with "large volatility" prompts. Experimental results in Table\ref{tab:ds_task_eval} and Figure\ref{fig:ds_task} show that as the amount of simulation data increases, the model benefits from training in a more comprehensive and reasonable market environment, leading to significantly improved performance on unseen real market conditions. This enhancement results in higher annualized returns, reduced drawdowns, and an improved Sharpe ratio. Moreover, simulation high volatility also contributes to increase portfolio returns, particularly during periods of high market fluctuations.

\begin{table}[t]
\centering
\caption{Evaluation metric of downstream models.}
\vspace{-2mm}
\resizebox{\columnwidth}{!}{
\begin{tabular}{lcccc}
\toprule
\small Training Set  & \small Annualized Return & \small Max Drawdown & \small Sharpe Ratio \\
\midrule
\small w/o sim.  & {20.67\%}  & {-0.029} & {5.13}   \\
 \small  1$\times$sim. & {24.10\%}  & {-0.028} & {5.50}   \\
 \small  5$\times$sim.   & {25.88\%}  & {-0.025} & {5.61}   \\
\small 10$\times$sim.   & {26.08\%}  & {-0.025} & {5.60}  \\
\small 10$\times$vol-sim.   & {27.39\%}  & {-0.026} & {5.71}  \\
\small 10$\times$sim.+10$\times$vol-sim.   & {28.46\%}  & {-0.022} & {6.32}   \\

\bottomrule
\end{tabular}\label{tab:ds_task_eval}}
\end{table}

\subsection{Ablation Study}
\paragraph{Retrieval method and Generation Model} 
To demonstrate that the design of similar sequence retrieval and conditional diffusion generation are both essential for the generative model, we will compare the performance differences between different settings.

In this study, we test the generation scenario using daily generation frequency, experimenting with different retrieval strategies and synthetic model. The results demonstrate that our FWT requires both correlation retrieval and diffusion-based generative model to simulate reliable data, while retrieve demonstrate higher contributions.

\begin{table}[t]
\centering
\caption{Ablation studies on retrieval methods and model.}
\label{tab:ablation}
\resizebox{\columnwidth}{!}{
\begin{tabular}{lccccccccc}
\toprule
Retrieval Method &  Model & Market Ranking & Correlation    \\
\midrule
\multirow{4}{*}{Ex-return corr}     & FWT   &  99.71\%  & 0.645   \\
                         & Transformer     &  92.06\%  & 0.385  \\
                                                  & GAN     &  96.42\%  & 0.411  \\

                         & Linear     &  83.29\%  & 0.226  \\
\midrule
\multirow{4}{*}{DTW}     & FWT   &  97.48\%  & 0.483  \\
                         & Transformer     &  92.15\%  & 0.337  \\
                                                  & GAN     &  92.06\%  & 0.460  \\

                         & Linear     &  75.92\%  & 0.181  \\
\midrule
\multirow{4}{*}{Random sampling}       & FWT  & 69.81\%  & 0.240       \\
                         & Transformer    &  59.62\%  & 0.203 \\
                                                  & GAN     &  51.04\%  & 0.238  \\

                         & Linear    &  59.73\%  & 0.198  \\
\midrule
\multirow{4}{*}{W/o retrieve}     & FWT     &  51.42\%  & 0.052     \\
                         & Transformer    &  51.26\%  & 0.047 \\
                                    & GAN     &  48.57\%  & 0.006  \\

                         & Linear    &  47.98\%  & 0.003  \\
\bottomrule

\end{tabular}}

\end{table}

\begin{figure}[th]
  \centering
  \includegraphics[width=.7\linewidth]{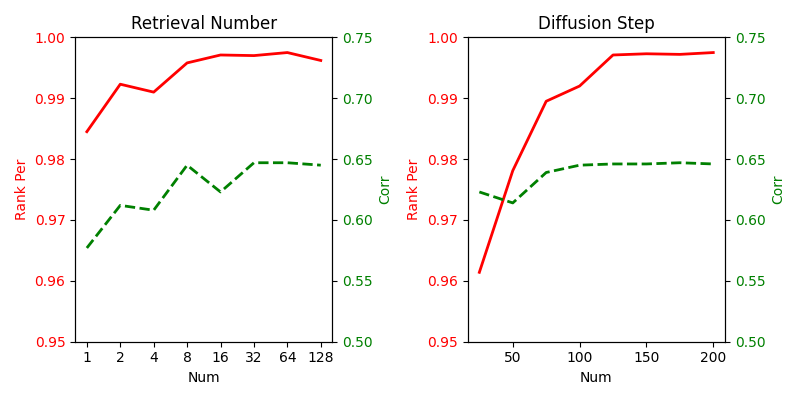}
  \vspace {-1em}
  \caption{Hyperparameter Sensitivity: Retrieval Number and Diffusion Step.}
  \label{fig:ablation}
\end{figure}

\paragraph{Hyperparameter Sensitivity} 
FWT has demonstrated good generative capability across various market conditions. Regarding the sensitivity to hyperparameters, we will test the most important parameters: the number of retrievals and the number of diffusion steps. In the previous experiments, we retrieve 16 similar stocks as the condition, with 100 diffusion steps. We compare the performance on two key metrics under different hyperparameters in Figure\ref{fig:ablation}. The conclusion is that the model is not very sensitive to these two parameters and consistently maintains stable generative capability.

\section{Conclusion}

In this work, we introduced FWT, a high-precision financial market simulator that enhances the financial time-series data generation through sequences retrieval augmentation. We designed a novel model structure, proposed and validated indicators for evaluating the quality of generation, tested FWT's performance across various market frequencies and conditions, and assessed the quality of cross-frequency generation. Additionally, we conducted tests in multiple market environments and cross-market pattern learning. We have also implemented "what-if" generation to simulate trends under customized future market assumptions. In downstream tasks, FWT provides risk testing and incremental training environments for multiple quantitative models, improving prediction performance and robustness to a great extent.

\bibliographystyle{ACM-Reference-Format}
\bibliography{main}

\appendix

\end{document}